# Tailored Security: Building Non-Repudiable Security Service Level Agreements

Takeshi Takahashi , Joona Kannisto , Jarmo Harju , Seppo Heikkinen , Bilhanan Silverajan , Marko Helenius , Shin'ichiro Matsuo

*The security features of current digital services are mostly defined and dictated by the service provider. A user can always decline to use a service whose terms do not fulfil the expected criteria, but in many cases even a simple negotiation might result in a more satisfying outcome. This article aims at making the building of non-repudiable security service level agreements between a user and a service provider more feasible. The proposed mechanism provides a means to describe security requirements and capabilities in different dimensions, from overall targets and risks to technical specifications, and it also helps in translating between the dimensions. A negotiation protocol and a decision algorithm are then used to let the parties agree upon the security features used in the service. This article demonstrates the feasibility and usability of the mechanism by describing its usage scenario and proof-of-concept implementation, and analyzes its non-repudiability and security aspects.*

## Introduction

The number of online services, including cloud computing services, has grown drastically. This trend has been accelerated by people's adoption of mobile communication devices into their daily lives. These have now become indispensable tools as they are used to access online services such as social networking, chat tools, email, online banking, and ticket reservation agents. Consequently, the number of online transactions has also grown drastically, and this has led to an increase of security incidents that threaten security and privacy in cyber-society.

Mechanisms to protect users against these risks are needed. Though various techniques exist for providing security for users, high costs and potential impairment of service usability make implementing all of them unrealistic. Users and service providers thus wish to strike a balance between security and usability, but building unique criteria for defining this balance is an impractical approach since the needed security measures differ depending on users' circumstances and environments. The balance needs to be defined for each user and each occasion of service usage.

Several hurdles stand in the way of realizing such a balance. First, users' security requirements need to be described in a machine-readable manner. Structured format for that is needed, but that is not enough. It is extremely difficult for ordinary users with very little technical knowledge to identify needed security techniques, and this is the case with many users accessing Internet services via mobile devices. They are thus unable to identify needed techniques even if they are aware of security risks. We therefore need techniques that let such non-technical users specify security requirements. Second, we need an automatic negotiation scheme to agree upon an SSLA. Users are currently allowed to choose only to agree or disagree with the security policies proposed by service providers and have no means to negotiate needed security levels and/or techniques with the provider. Even if the provider wishes to negotiate with users, manual negotiation is impractical in terms of cost, so an automatic negotiation scheme is needed. Third, the agreement needs to be non-repudiable for the negotiation parties. No matter how amenably a user and service provider agree upon the needed level of security, security incidents may eventually occur. If an incident stems from a violation of the agreement, one party may wish to sue the other. Thus, the agreed level of security needs to be non-repudiable for both parties.

To address the above issues and realize a balance between usability and security, this article introduces a mechanism that builds a non-repudiable security service level agreement (SSLA). An SSLA is the information on a service's security level agreed to by a user and service provider. The mechanism provides two techniques – security expression and translation. Security expression technique provides a means to describe security requirements and capabilities in multiple dimensions and in a machine-readable format. Translation technique provides a means to translate such information among different dimensions, which enables users with little technical knowledge to express security requirements with non-technical vocabulary and translate them into technical vocabulary. These techniques are used by the mechanism's negotiation protocol that enables the user and service provider to negotiate and agree to an SSLA. This enables service providers providing only yes-or-no choices to users on certain security policies to interact with users to set up agreeable level of security policies. The negotiation generates a non-repudiable SSLA by using cryptographic identities and digital signatures.

This article also demonstrates the usability and feasibility of the proposed mechanism by providing a usage scenario and proof-of-concept implementation, and then analyzes the mechanism.

## Key Challenges

To build a mechanism that generates a non-repudiable SSLA, a range of issues need to be considered. There are related works for each of the issues, and they are good basis for considering the mechanism.

### Machine-Readable Security Expression

Security requirements need to be expressed in a machine-readable manner in order to negotiate the SSLA over the Internet. One approach for this is using a structured language, such as XML or JSON, along with its schema. The WS-Agreement specification [1] provides an XML schema for specifying an agreement and the language in which advertising of service capabilities are performed. While it is intended for making business-based agreements, it should also be possible to use it for defining security-specific agreements, based on the extensibility of XML. Policy languages may also be used to express different types of requirements of the parties. WS-Policy [2] is one such language and it can be used along with other WS-* specifications – namely WS-Security [3], WS-Trust [4], and WS-SecureConversation [5] – to express the required security functionality for the established communication. These solutions, however, incur considerable overhead, as with many other XML-based solutions.

Even if we have a means to express security requirements in a machine-readable manner, several issues still exist. First, it is difficult to prepare a single schema for describing user requirements. Users with differing level of technical knowledge may wish to specify security requirements with different vocabularies or in different viewpoints. Users and service providers also have different viewpoints, so users wish to express security in terms of non-technical terms to broaden the scope of protection, while service providers do this in terms of specific technical terms to minimize the scope of responsibility. Second, allowing free text strings may hinder the automation of SSLA processing and negotiation. The proposed mechanism thus takes an approach that allows multiple dimensions of such a description, prohibits free text strings, and provides a means to translate between the dimensions.

### Negotiation and Agreement-Building Protocol

Negotiation needs to take place in order for two parties to agree on an SSLA. There is a good amount of existing research regarding the creation of SLAs. Quality of service (QoS) is one prominent application domain, and Hasselmayer et al. [6] suggest an architecture in the spirit of Web Services (WS) and apply grid technologies for the implementation. This basically provides discovery mechanisms for the multitude of service providers and XML templates for the presentation of the service offer – i.e., the SLA to be formed – which the client can accept or refuse. It also uses the WS-Agreement specification [1] for the actual protocol and the language in which the advertising of service capabilities is conducted.

Another approach for agreements is key exchange protocols, such as TLS [7], IKEv2 [8], and HIP [9]. Key exchange protocols provide a means to agree upon actual communication security parameters, and can also be modified to exchange the SLA type of data. Alternatively, they can be made to support the above XML-based specifications. HIP provides flexible mechanisms for protocol additions, but it is not a widespread protocol and has its own constraints regarding the amount of exchanged data. IKEv2 allows the possibility of defining new security associations, but as with TLS the original specifications still need to be modified. TLS, however, enjoys the most widespread deployment.

Based on the above works, we have developed a negotiation protocol that can cope with the vocabulary gap between a user and service provider.

### Non-repudiable Proof of Agreement

Non-repudiability in the SSLA negotiation context means that, at a minimum, it should be impossible for either party to refute the negotiation result. This may also concern the negotiation process itself in that any statement made during the negotiation cannot be refuted afterward. For the sake of fairness, both parties should retain the same amount of proof in order to avoid situations where one party holds a preponderance of proof [10]. The proposed negotiation protocol addresses the above issue by producing non-repudiable proofs with cryptographic identities and digital signatures.

## Architecture

This section introduces the architecture of the proposed mechanism and elaborates on the roles needed by the architecture and on the process.

### Roles

The proposed mechanism defines three roles – User, Service Provider (SP), and Knowledge Base (KB). **User** wishes to use an online service. It is aware of the high-level security requirements for the use of such services and activates the negotiation of an SSLA with the SP. The requirements differ depending on the target online service and any other circumstances of the user. Note that it can also express security capabilities if needed. **Service Provider** (SP) provides online services to Users. It knows what security measures it can provide and which user requirements it may satisfy by consulting the KB. Note that it can also express security requirements if needed. **Knowledge Base** (KB) provides knowledge concerning security upon request. It administrates and maintains dictionaries of security expressions in differing dimensions and translation tables between the dimensions.

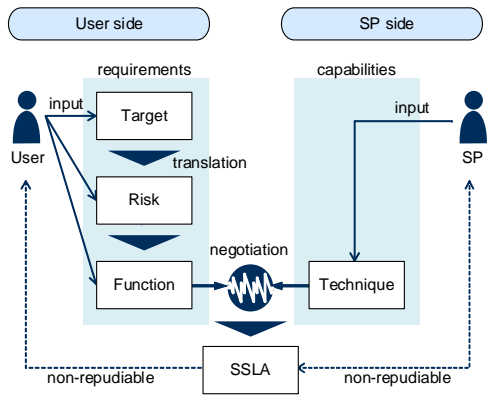

*Figure 1* Process Overview

## Process Overview

Figure 1 gives a simplified overview of the proposed mechanism, which provides a means for users to describe security requirements from various dimensions. It translates such requirements in various dimensions into requirements in a single dimension. It then negotiates the requirements and capabilities of both of the user and service provider and agrees upon the needed security requirements of the service, i.e., the SSLA. The sequence of the procedure is conducted automatically without manual intervention. The SSLA is also built so that it becomes non-repudiable.

## Fundamental Techniques

Several techniques are needed to realize the architecture. Indeed, security requirements and capabilities need to be described in a machine-readable manner and translated into arbitrary dimensions in order for a User and SP to negotiate a non-repudiable SLA. This section introduces four such techniques.

## Security Expression

An **SSLA** is the information on the agreed to security level between a User and SP. In order to build an SSLA, the proposed mechanism defines two types of security information: security requirement and capability. **Security requirement** is the information on what kind of security or security measure one party requires while **capability** is the information on what kind of security or security measures one party can provide. An SSLA is built through matching and negotiation of security requirements and capabilities of both a User and SP. The vocabulary for describing such information is stored inside dictionaries, which are stored inside KBs. To facilitate machine processing the proposed mechanism minimizes the room for free text input by assigning unique identifiers for each vocabulary item. The identifiers are expressed as Object Identifiers (OIDs) [11]

The vocabulary that Users and SPs wish to use differs. To facilitate the expression of security requirements and capabilities for various users, the mechanism defines four dimensions of such information – Target, Risk, Function, and Technique – and provides separate types of dictionaries. **Target** dimension expresses them from the standpoint of what to protect by specifying the target of the protection, such as user's "personal information." The vocabulary in this dimension is stored in the "Target" dictionary. **Risk** dimension expresses them from the standpoint of what types of risks to avoid by specifying risks that need to be avoided, such as the risk of "network sniffing." The vocabulary in this dimension is stored in the "Risk" dictionary. **Function** dimension expresses them from the standpoint of what types of functions need to implemented by specifying needed security functions, such as "user data encryption" and "user authentication." The vocabulary in this dimension is stored in the "Function" dictionary. **Technique** dimension expresses them from the standpoint of which technique an SP needs to implement by specifying needed security techniques, such as "AES" and "SHA." The vocabulary in this dimension is stored in the "Technique" dictionary.

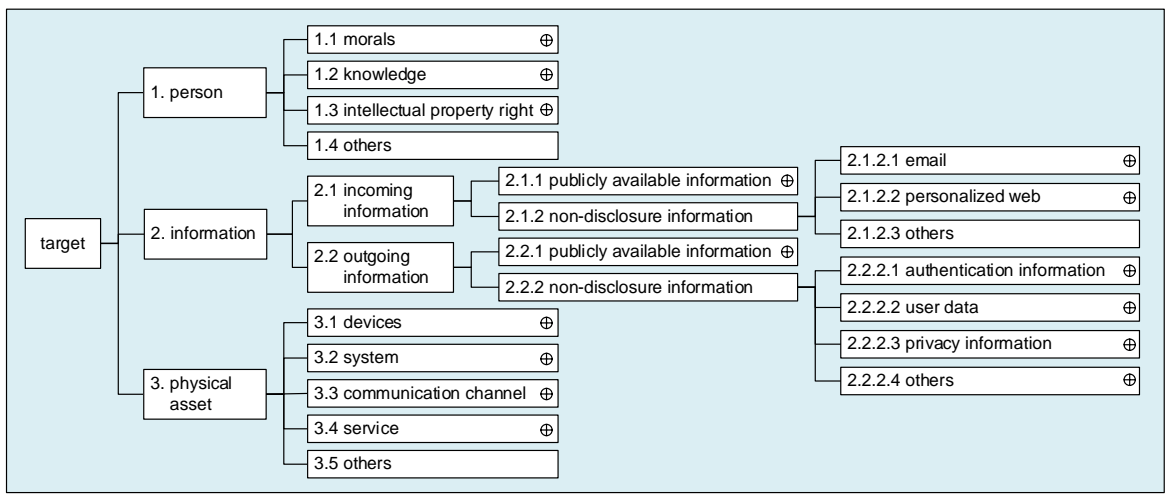

*Figure 2* Excerpt of Example Target Dictionary

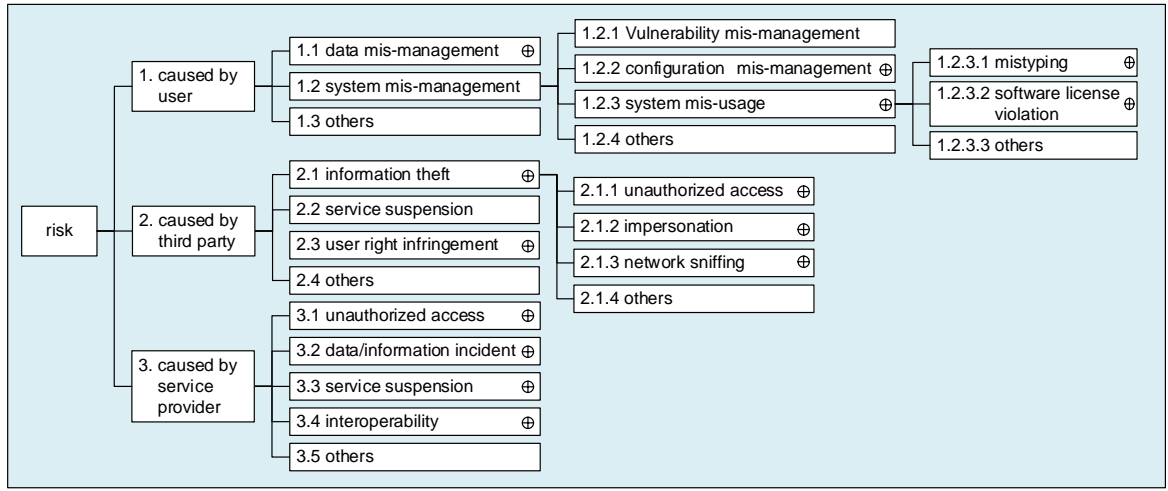

*Figure 3* Excerpt of Example Risk Dictionary

Vocabularies in these dictionaries of different dimensions begin with TARGET, RISK, FUNCTION, and TECHNIQUE – each assigned unique OID arcs. Since users and service providers typically have multiple security requirements and capabilities, the security requirement, capability, and SSLA are indeed expressed as the list of these OIDs. Each KB may have its own dictionaries. To demonstrate the construction of these dictionaries, Figures 2 and 3 show excerpts of Target and Risk dictionary examples. The numbers in the figures are preceded by the OID arcs specifying the types of the dictionaries: TARGET, RISK, FUNCTION, and TECHNIQUE.

We sometimes wish to express security by using vocabulary of multiple dimensions. We can combine arbitrary OIDs in differing dimensions by using colons. For instance, we wish to specify a function to cope with a specific risk. In this case, we can concatenate OIDs of function and risk by using a colon between them, e.g., "Risk.1.1.2:Function.19.12.2." This feature is useful especially when counter-proposing an SSLA.

### Translation

By providing differing dimensions of security expressions, users can express security from various dimensions so they do not fail to identify important security requirements. Though various types of users may enjoy this feature, a technique to translate information from one dimension into another is needed in order to process information of differing dimensions.

*Table 1* Excerpts of example translation tables

(a) [target, risk] table

| target | Risk |
|---|---|
| Target.1.1.1 | Risk.2.3.4 |
|  | Risk.3.2.3 |
| Target.1.1.2 | Risk.2.2.5 |
|  | Risk.2.3.2 |
|  | Risk.3.1.3 |

(b) [risk, function] table

| risk | function |
|---|---|
| Risk.1.1.1 | Function.12.1.3 |
|  | Function.17 |
|  | Function.23.3 |
| Risk.1.1.2 | Function.15 |
|  | Function.19.12.2 |

The proposed mechanism provides a technique for such translation. It looks up translation tables, which map an OID of one dimension to corresponding OIDs of differing dimensions. The tables are stored inside KBs.

Three types of translation tables exist: [target, risk], [risk, function], and [function, technique]. The tables basically have two columns. Typically an OID of one column corresponds to multiple OIDs of another column. The [risk, function] translation table, for instance, have two columns – one for OIDs of risks and another for the ones of functions. The table has multiple entries for one risk with differing functions since several functions are required to cope with a certain risk. Table 1 shows excerpts of example translation tables.

### Negotiation protocol

The security expression and translation techniques are useful for a User and SP negotiating an SSLA since Users often wish to express security requirements with non-technical vocabulary while SPs often wish to agree upon an SSLA using specific technical vocabulary. The mechanism defines two types of communication: KB lookup and SSLA negotiation.

**KB lookup** is performed to translate requirements and capabilities of various dimensions. It uses Translation-request and -reply messages. Users and SPs can query KBs to translate requirements and capabilities in one dimension into another by sending a Translation-request that includes them. The KBs reply with a Translation-reply message that contains the translated requirements and capabilities.

**SSLA negotiation** takes place to build an SSLA between two parties. SSLA-proposal and -confirmation messages are used. One of the parties sends an SSLA-proposal message that contains its requirements and capabilities, and the other party replies with an SSLA-confirmation message if it agrees to satisfy the requirements. If it disagrees, it can generate new requirements considering the received capability and reply with new SSLA-proposal message. This message exchange continues until one party replies with an SSLA-confirmation message or cancels the negotiation. The negotiation is terminated when the SSLA-confirmation message has reached its recipient. The list of agreed requirements becomes an SSLA.

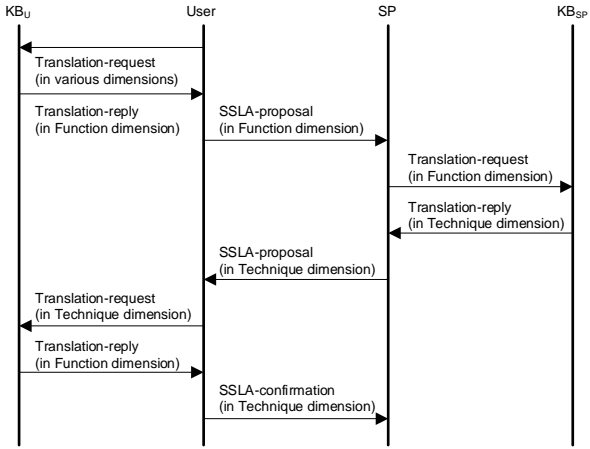

**Figure 4** One-round Negotiation Sequence

Though multi-round negotiation may take place, for simplicity, Figure 4 describes simple one-round negotiation. Note that $KB_U$ and $KB_{SP}$ are the KBs that the User and SP trust respectively. Prior to initiating negotiation, a User communicates with $KB_U$ to translate security requirements in differing dimensions into the Function dimension. The User then sends an SSLA-proposal message that contains the requirements and optionally the URI of $KB_U$. Upon receiving the message, the SP considers whether it can satisfy the proposed requirements and wishes to specify requirements in Technique dimension rather than Function dimension so that the requirements can be more specific on its responsibility. It thus communicates with $KB_{SP}$ to know the techniques that satisfy the User's requirements. It then constructs a new SSLA-proposal message that contains the list of techniques and optionally the URL of $KB_{SP}$. Upon receiving the message, the User checks whether the techniques can satisfy its requirements by communicating with $KB_U$. Once that is confirmed, it sends an SSLA-confirmation message to the SP. Note that in this example the security requirement sent from the User at the beginning of the negotiation is described in the Function dimension while the SSLA that is agreed to at the end is described in the Technique dimension.

The messages of the negotiation protocol use cryptographic identities and digital signatures to make the resultant SSLA non-repudiable. The SSLA-proposal message contains the list of requirements and capabilities, the identities of the negotiating parties, negotiation ID, nonce, timestamps, signature, and proof-of-work token. The identities should be linked to the public keys of the participants, such as through a one-way hash function of the public key or a certificate. The negotiation ID, which remains the same for the entire negotiation, and nonce protect the sender from replay attacks. The message is signed with the private key corresponding to the sender's public key. The proof-of-work token was generated by using a hashcash stamp [12]. The generation of the stamp requires the sender to commit a parametrizable amount of computation work, and thus contributes to minimizing the risk of DoS attacks. The SSLA-confirmation message contains the contents of the SSLA-proposal message. It also contains the sender's signature and nonce as well, as with the SSLA-proposal message.

When the SSLA-confirmation message reaches its recipient, both sender and recipient have the list of requirements, i.e., the SSLA, with their signatures. In this way, both parties hold the same information that is usable as a proof.

### Decision Algorithm

When an SSLA-proposal message is sent, the recipient needs to decide whether to accept it, create a counterproposal, or abort the negotiation, as we have seen above. This decision could be performed by manual operation, but this should be automated in order for an SP to handle a high volume of Users.

One simple approach is that the recipient accepts the proposal if it has sufficient capability, as described in Algorithm 1, where *req* represents an entry of the requirements, *caps* represents the capabilities, the *dimension* subroutine returns the dimension of its argument, and the *translate* subroutine translates its argument into one of the dimensions, and that is performed by the above KB lookup procedure. The algorithm returns either *1* or *0*, where *1* means that the *caps* can satisfy the *req* and thus are sufficient as SSLA entries. The algorithm needs to be executed for each entry of the requirements.

Algorithm 1 begins with a branch on the condition. If *req* is described in the Technique dimension, it is simply compared with the list of capabilities. If it matches one entry of *caps*, the algorithm returns *1*. If *req* is described in a non-Technique dimension, it is translated into *translated-reqs* that are described in the Function dimension. Note that the *translation*

```
if {dimension(req)=Technique} then
    if {req ∈ caps} then
        return 1
    else
        return 0
    end if
else
    translated-reqs <- Translate(req, Function)
    for all entry in translated-reqs do
        if entry ∉ Translate(caps, Function) then
            return 0
        end if
    end for
    return 1
end if
```

**Algorithm 1** Simple decision algorithm

subroutine can convert one requirement into several requirements in a different dimension. Each entry of the *translated-reqs* is compared with the list of capabilities that are translated from the Technique dimension to the Function dimension. If all entries of the *translated-reqs* match the capabilities, the algorithm returns *1*.

In cases where some or none of the requirements are satisfied by the *caps*, the recipient cannot accept the SSLA proposal. It can then counter-propose another SSLA or cancel the negotiation. If the negotiation is cancelled, the sender could, for instance, change *req* and redo the procedure.

## Discussion and Analysis

This section demonstrates the mechanism's usability and feasibility by discussing its usage scenario and by introducing our proof-of-concept implementation respectively. It also analyses the mechanism's non-repudiability and security features.

### *Usage Scenario*

One usage scenario of the mechanism is that a user of a mobile device wishes to use a hotspot service provided by a service provider. The user stipulates its security requirements and capabilities, negotiates with the service provider, and finally receives the service. The detailed process is elaborated below.

The user is required to input his credit card information for settlement in order to use the service. Nevertheless, the user fears that the information can be leaked or mistreated, so he chooses the risk of "network sniffing" and "data mismanagement" from the Risk dictionary and stipulates their OIDs. Prior to the negotiation, he communicates with his trusted KB that translates them, and receives the translated OIDs that indicate that "stored data encryption," "communication data encryption," "authentication," and "complete data removal after usage" are the needed functions. Now he is ready to begin an SSLA negotiation with the service provider and sends the OIDs of his security requirements to the provider. At the same time, he sends OIDs of his capabilities that include the "S/Key" authentication capability.

Upon receiving the OIDs, the provider considers whether it can satisfy the requirements. The requirements are somewhat vague for the provider, thus it wishes to agree upon the SSLA in more concrete dimension. It then asks its trusted KB, which replies with the OIDs of techniques that can satisfy the requirements in case of the first three requirements. The provider then looks up the capabilities it received from the user since some techniques requires the user to have some capabilities. In case of the third requirement, for instance, the provider decides to use the "S/Key" technique since it is included in the list of techniques suggested by the KB and is also included in the user's capability list. Regarding the fourth requirement "complete data removal after usage," the KB replies without any translation, meaning that the requirement is already concrete enough and the provider can satisfy that. Now the provider is ready to counter-propose a more concrete SSLA that satisfies the user's requirements.

The provider responds to the user with the list of techniques that are concatenated with the original requirements with colons. Upon receiving the answers, the user checks whether the techniques can satisfy his requirements by requesting translation from the trusted KB. Once he has confirmed that the proposed techniques can satisfy the requirements, he sends the confirmation message to the provider. In this case, the list of techniques becomes the SSLA. Now the user is ready to use the service.

### *Proof-of-Concept Implementation*

A proof-of-concept implementation was made to verify the feasibility of the proposed mechanism. It was implemented as a REST-based service with Java using JAX-RS API [13] and BouncyCastle cryptographic [14] libraries. The requirements, capabilities, and SSLA contents were described as JSON objects that contain the list of security expression identifiers in OID format.

The resource identifiers in the service were derived from the hashcash stamp, which forms a unique identifier for each negotiation if it includes the MAC, identities of both negotiation parties, and a random nonce. The implementation used hashcash's extension field and included these items of information by defining its own information structure.

The hashcash protocol needs to know the puzzle's difficulty, and the timestamp's granularity before the connection establishment. Therefore, some values should be used as the default. The implementation sets a default stamp difficulty at a 12 bit collision. It took computing resources in the order of 10-30 milliseconds (depending on the used SHA-1 implementation) using a Python-based implementation for generating a hashcash stamp [15].

The service was able to generate a non-repudiable SSLA through negotiation, and that demonstrates the feasibility of the mechanism. Note that the source code of the implementation is available online [16].

### *Non-repudiability consideration*

The mechanism should be able to resolve any dispute about a committed transaction solely based on the signatures of the transaction participants, and it should not have to rely on the existence of a third party for dispute resolution. It is obvious that our proposed mechanism can achieve that since the negotiation is taken place directly between a User and SP without any third party, and their signatures verify the contents of the SSLA. It indeed imitates the traditional process of negotiating about the contents of the agreement and finally mutually signing two copies of a paper that defines the contract, one for each party.

It is desirable that the fairness of negotiation is guaranteed. That is, none of the parties should be able to gain any advantage for having some partial evidence. Achieving the fairness is extremely complicated in the digital world, and the proposed mechanism is unable to guarantee that entirely. For instance, each party can propose a contract, but as one of them initiates the negotiation offering, the other might be considered to have a slight advantage. In particular, one party ends up in the possession of a signed agreement before the other.

Sending a signed SSLA proposal to the SP may not normally be of any concern to the User, since the User might not start using the service before having received a signed SSLA from the SP. We may, however, imagine some rare cases where a mismatch in the synchronization of a series of negotiation attempts would make it possible for one party to exploit a misunderstanding of the other party on the security level applied on the service.

There are various solutions for this synchronization problem, e.g., the user of the service may require an explicit reference to the applied SSLA, or it may be required that the parties always follow the latest agreed SSLA in the transactions.

*Security Considerations*

The SSLA negotiation imposes non-lightweight tasks to the communication parties. Thus it needs to consider DoS resistance feature. DoS attacks make one or more resources that a service offers to its legitimate users inaccessible to them. The attacker in DoS attacks commonly depletes finite resources, such as computational power, communication bandwidth, and the ability to maintain state. An attacker can, as a basic rule, cause a DoS attack on any service if the attack is distributed and powerful enough, but shortcomings and weaknesses in protocols and services can make attackers' tasks easier. For instance, the target service usually cannot discern the attacker from the legitimate users, and the attacker can make an overwhelming amount of normal requests consuming all the resources. Therefore, DoS resistance usually means that an attack is not possible without considerable effort on the attacker's part. If the attacker has to commit more resources than the defending party, the economies of the situation support the defender.

Protocols basically need proof that the other party is committed to the transaction before any heavy work or any types of resource allocation can be performed in order to be DoS-resistant. This proof also needs to be verified with little effort. Different kinds of proofs of commitment can be made; some are computational while others may rely on other factors, for instance on the communication party's ability to communicate and maintain a state.

The proposed protocol requires the SP to perform multiple computations including matching between requirements and capabilities upon receiving the first message of the protocol from the User. It thus needs to consider a DoS resistance feature. It factors that in and uses a hashcash stamp [12]. This constitutes a proof-of-work that takes a parameterizable amount of work to compute for the sender. The recipient can efficiently verify the received hashcash stamps. This induces a performance penalty on the initiator, hence mitigating DoS from the server's perspective. The use of hashcash in the protocol makes it resistant against DoS, and that DoS-resistant feature is needed for the protocol since creating the SSLA is a non-trivial effort for the SP.

A signature in the user's first packet would also be useful for DoS-prevention purposes, at least if RSA signatures are used. There is then the possibility of even removing the hashcash stamp. Indeed, since RSA signature verification is much faster than signature generation, with the usual public exponents a valid signature could serve as a proof of commitment. While checking RSA signatures is not as fast as hashcash puzzle validation, the signature would also serve the other purposes: as a proof of the ownership of the public key, and as a signature for the final SSLA.

There are many security aspects apart from DoS resistance feature. Though they are not particularly mentioned in this article, many existing techniques can be used to reinforce its security feature. For instance, the use of message authentication codes and signatures guarantees that our mechanism protects the parties from typical security threats, such as message interception and modification, and replay attacks. It can also prevent the message sender from denying the sending of a message. The mechanism can also ensure the confidentiality of transactions, whenever the nature of the agreement requires confidentiality, by implementing various schemes, such as JSON Web Encryption (JWE) [17].

## Conclusion and Future Works

The proposed mechanism builds non-repudiable SSLA by using its security expression technique, translation technique, negotiation protocol, and decision algorithm. The usage scenario and the proof-of-concept implementation clarified the usability and feasibility of the mechanism, and the discussion section analysed its non-repudiability and DoS resistance features. One major issue for further work is how to automatically identify security requirements and capabilities of users. This is especially an important issue for users of mobile tablets since their user interface including display and input devices are limited. Moreover, the requirements and capabilities may change dynamically depending on circumstances including the hotspot services the users uses and the remaining amount of battery. Through continuing this research, we believe we contribute to realize tailored security that is secure and usable.

## Author Information

*Takeshi Takahashi (takeshi_takahashi@ieee.org):* received a Ph.D. in telecommunication from Waseda University in 2005. He worked for the Tampere University of Technology in 2002-2004 as a researcher and Roland Berger Ltd. in 2005-2009 as a business consultant. Since 2009, he has been working for the National Institute of Information and Communications Technology and is currently a senior researcher there. His research interests include Internet security and network protocols.

*Joona Kannisto* (joona.kannisto@tut.fi): received an M.Sc from Tampere University of Technology in 2011. He is a researcher and Ph.D. candidate at the university and has been involved in security related research ever since. His research interests include secure protocols, usable security, and reputation and trust management.

*Jarmo Harju* (jarmo.harju@tut.fi): received a Ph.D. in mathematics from the University of Helsinki in 1984. In 1985 - 1989 he was a senior researcher at the Technical Research Center of Finland. In 1989 - 1995 he was a professor at Lappeenranta University of Technology. Since 1996 he has been a professor of telecommunications at Tampere University of Technology, where he is leading a research group concentrating on network architectures and network security.

*Seppo Heikkinen* (seppo.heikkinen@tut.fi): received a Ph.D. from Tampere University of Technology in 2011. He worked for Finnish telecom operators in 1997-2005 and in 2006-2012 as a researcher at the university concentrating on network security. He currently works for Nixu Ltd. as a security consultant with emphasis on PCI DSS compliance. His research interests include payment security, future network security architectures, identity-based security, and non-repudiation mechanisms.

*Bilhanan Silverajan* (bilhanan.silverajan@tut.fi): received an M.Sc from Lappeenranta University of Technology in 1998. He is completing his PhD at Tampere University of Technology. Since 1998 he has been actively involved with scientific research and project management of national as well as EU-wide collaboration projects for future Internet architectures, network and device mobility, pervasive middleware, home networks, and ubiquitous services, as well as the Internet of Things.

*Marko Helenius* (marko.t.helenius@tut.fi): received a Ph.D from University of Tampere in 2002. He joined Tampere University of Technology in 2008 as a senior researcher. His current research interests include usable security, secure programming and cloud security.

*Shin'ichiro Matsuo* (smatsuo@nict.go.jp): received a D.E. degree in 2003 from Tokyo Institute of Technology. In 1996-2009, he worked for NTT DATA Corporation. Since 2009 he has been working for the National Institute of Information and Communications Technology and is currently a director there. His research interests include information security and cryptographic protocols.